\documentclass{article}
\usepackage{spconf,amsmath,graphicx,hyperref}
\usepackage{setspace}

\title{Two-stage Audio-Visual Target Speaker Extraction System for Real-Time Processing On Edge Device}
\name{Zixuan Li$^{1}$, Xueliang Zhang$^{1}$\sthanks{Corresponding author.}, Lei Miao$^{2}$, Zhipeng Yan$^{2}$, Ying Sun$^{2}$, Chong Zhu$^{2}$}
\address{$^{1}$College of Computer Science, Inner Mongolia University, China\\
$^{2}$Lenovo, China\\
\small \texttt{cslzx@mail.imu.edu.cn, cszxl@imu.edu.cn, miaolei1@lenovo.com, yanzp@lenovo.com}
}
\usepackage{amssymb}
\usepackage{booktabs}
\usepackage{utfsym}
\begin{document}
\ninept
\setstretch{0.87}
\maketitle
\begin{abstract}
Audio-Visual Target Speaker Extraction (AVTSE) aims to isolate a target speaker's voice from multi-speaker mixtures by leveraging visual cues. However, the practical deployment of existing AVTSE methods is often hindered by poor generalization, high computational complexity, and non-causal designs. To address these issues, we propose 2S-AVTSE, a novel two-stage system built on an audio-visual decoupling strategy. This approach uniquely eliminates the need for synchronized audio-visual training data, enhancing its applicability in real world scenarios. The first stage uses a compact visual network to perform voice activity detection (VAD) by analyzing visual cues only. Its output VAD then guides a second-stage audio network to extract the target speech. With a computational load of only 1.89 GMACs, our system exhibits superior generalization and robustness in realistic and cross-domain scenarios compared to end-to-end baselines. This design presents a practical and effective solution for real-world applications.
\end{abstract}
\begin{keywords}
Audiovisual System, audio-visual target speaker extraction, target speaker extraction, real-time system
\end{keywords}
\section{Introduction}
\label{sec:intro}

The target speaker extraction (TSE) system aims to isolate the voice of the target speaker in noisy environments with multiple interfering speakers. TSE systems leverage spatial, audio, visual, or semantic cues to separate target speech in complex acoustic environments, offering a practical solution to the cocktail party problem\cite{cherry1966human,vzmolikova2019speakerbeam}. 
Currently, most TSE systems rely on audio cues \cite{ju2023tea2, ju2023tea3, zhang2023real, lin2023focus}, where the audio cue is a pre-recorded reference speech of the target speaker, called the Anchor. 
However, audio cues can not reliably identify the target speaker, especially when different speakers have similar voice characteristics or when voice features are affected by health conditions. 
Additionally, such systems require the user to pre-record an Anchor, which is inconvenient in real-world applications.

Inspired by the human ability to integrate visual and auditory cues for robust speech perception in noisy, multi-speaker environments \cite{sumby1954visual,crosse2016eye}, a growing body of research has focused on audio-visual approaches to enhance speech signals \cite{sato2021multimodal,li2024audio,kalkhorani2024audiovisual,kalkhorani2024av}.
However, the practical deployment of these AVTSE systems for real-time applications, such as online conferencing, remains challenging. Several key limitations hinder their widespread adoption:

\textbf{1.Lack of Realism in Simulated Data:} A significant issue stems from the artificial nature of the training and evaluation data. First, widely-used datasets for AVTSE and speaker separation, such as LRS2-2mix, LRS3-2mix, and VoxCeleb2-2mix\cite{li2024audio}, are generated by mixing two speech signals with a constant 100\% temporal overlap, a scenario rarely encountered in practice. Compounding this issue, the mixtures are often created by the simple summation of source signals recorded under disparate acoustic conditions. Consequently, the model may learn to exploit these artificial differences in the acoustic environment as a spurious separation cue—one that is absent in any real-world recording where all speakers share the same space. 

\textbf{2.Poor Generalization due to Data Scarcity:} The prevalent end-to-end training paradigm for these AVTSE systems poses a significant generalization challenge. 
While end-to-end models can achieve high performance, their generalization capabilities are highly dependent on the availability of large-scale, diverse training corpora.
However, synchronized audio-visual data is considerably scarcer than its unimodal audio or visual counterparts. Consequently, resolving the poor generalization by simply scaling up the training corpora would be prohibitively expensive.

\textbf{3.Prohibitive Complexity and Non-Causal Architectures:} The high computational complexity of these models is often prohibitive for resource-constrained hardware. Moreover, their non-causal architectures render them fundamentally unsuitable for real-time processing.

To address the aforementioned challenges and facilitate the practical deployment of AVTSE, this paper introduces \textbf{2S-AVTSE}, a novel two-stage framework. The core innovation of our approach is a decoupled training strategy that does not require synchronized audio-visual data. This allows the system to benefit from vast corpora of high-quality, audio-only data and leverage realistic acoustic simulations via established techniques, such as the Image method\cite{allen1979image}. As a result, the proposed 2S-AVTSE achieves superior generalization while maintaining a lightweight computational costs, making it a viable solution for real-world applications. Our main contributions are summarized as follows:

1. We propose a novel two-stage, decoupled training paradigm for AVTSE that eliminates the need for synchronized audio-visual data. This framework performs visual voice activity detection (VVAD) and then uses its output to guide target speech extraction. We demonstrate that this strategy achieves superior generalization in realistic scenarios compared to conventional end-to-end approaches.

2. We introduce a significant simplification of the visual front-end by replacing complex lip-reading encoders with a highly efficient VVAD network (0.18 GMACs). To overcome data scarcity and class imbalance for its training, we innovatively leverage 3D talking portrait generation to create a large-scale, balanced dataset.

3. The complete 2S-AVTSE system is extremely lightweight, requiring only 1.36M parameters and 1.89 GMACs to effectively suppress both noise and interfering speakers. This high efficiency makes it practical for deployment on personal computers and other resource-constrained devices.

\begin{figure*}[htbp]
    \centering
    \includegraphics[width=\textwidth]{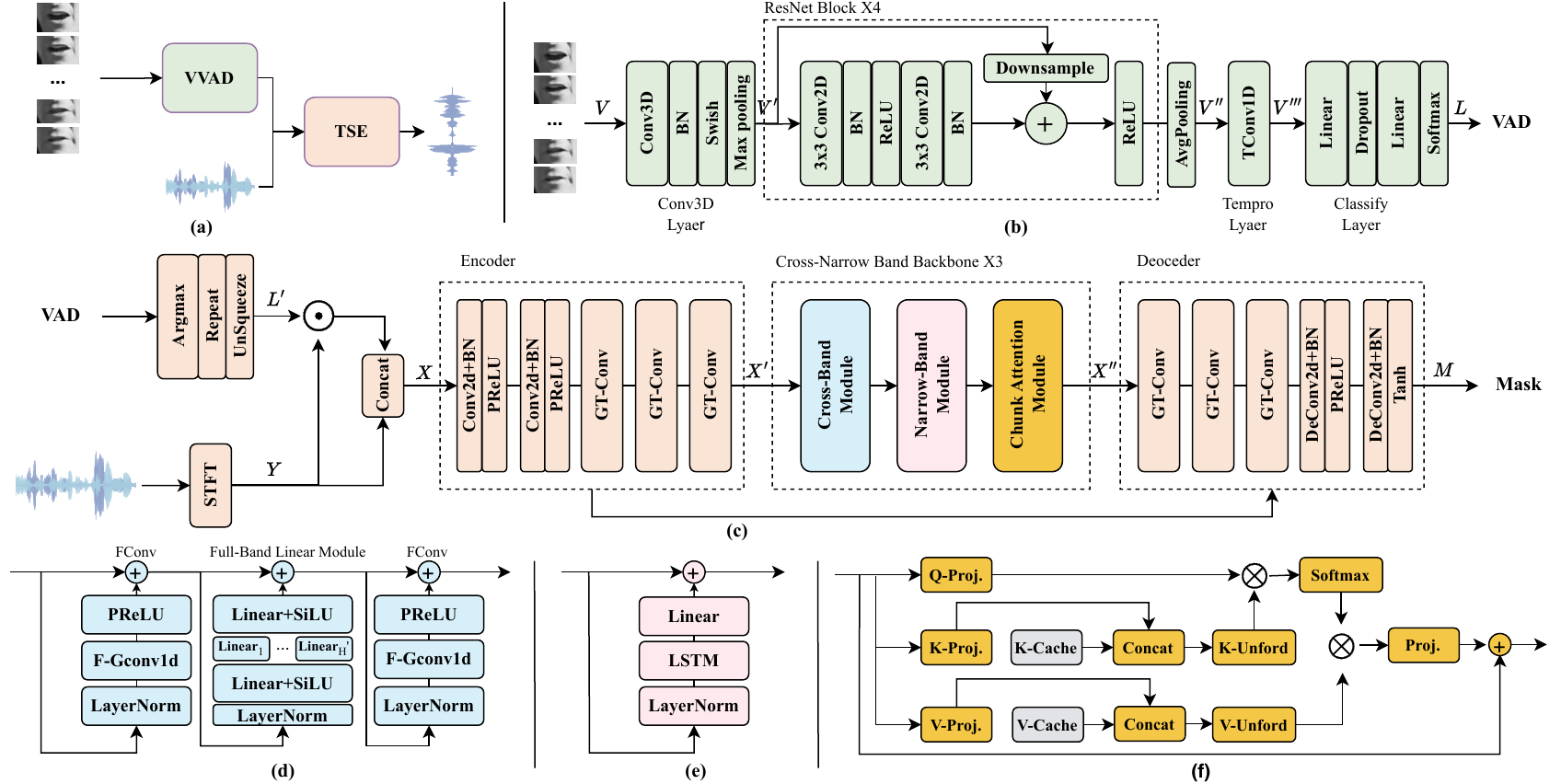}
    \caption{ overview of the 2S-AVTSE architecture along with detailed structures of its individual components. (a) Overview of 2S-AVTSE, VVAD module is ths first stage, TSE module is the second stage. (b) Overview of VVAD Module. (c) Overview of the TSE Module, where $\odot$ represents element-wise multiplication. (d) Overview of Cross-Band Module. (e) Overview of Narrow-Band Module. }
    \label{fig:big-fig}
\end{figure*}

\section{2S-AVTSE}
\label{sec:method}

The overall architecture of the proposed 2S-AVTSE system is illustrated in Figure \ref{fig:big-fig}(a). In the first stage, continuous lip video frames are processed by the visual voice activity detection (VVAD) module to determine the target speaker’s VAD. In the second stage, the VAD and the mixture’s complex spectrum are input to the TSE module, which integrates features from both modalities to estimate a complex ratio mask (CRM \cite{williamson2015complex}). The CRM is applied to the mixture’s complex spectrum, and the target speech is reconstructed using the inverse short-time Fourier transform (iSTFT).

\subsection{First Stage: Visual Voice Activity Detection}
We assume all videos are recorded at 25 frames per second, with the target speaker’s mouth region converted to a grayscale image of size 1 $\times$ 32 $\times$ 32. If multiple speakers appear in a frame, the speaker closest to the camera (i.e., with the largest lip region) is assumed to be the target. If no speaker is present, the lip region is represented as a zero matrix of the same size. The VVAD module processes input lip frames $V \in \mathbb{R}^{1 \times T_{v} \times 32 \times 32}$, where $T_{v}$ denotes the number of video frames, cropped and scaled from the video stream.

The VVAD module, shown in Figure \ref{fig:big-fig}(b), includes a Conv3D layer, four ResNet Blocks \cite{he2016deep}, a Temporal Layer, and a Classification Layer. The Conv3D layer captures spatiotemporal features using a 3D convolution with a kernel size of (5, 7, 7), a stride of (1, 2, 2), and 32 channels, followed by batch normalization (BN), an activation function, and max pooling (kernel size: (1, 3, 3), stride: (1, 2, 2)), producing $V^{\prime} \in \mathbb{R}^{32 \times T_{v} \times 8 \times 8}$. Next, four ResNet Blocks with 32, 48, 64, and 128 channels extract spatial features. Each block includes two 3$\times$3 convolutions with BN, activation functions, and Downsample for residual alignment. An average pooling layer reduces spatial dimensions to 1$\times$1, yielding global features $V^{\prime \prime} \in \mathbb{R}^{128 \times T_{v}}$ after reshaping. The Temporal Layer models temporal correlations between frames, focusing on mouth movements, using a 1D convolution with 32 channels and a kernel size of 5, resulting in $V^{\prime\prime\prime} \in \mathbb{R}^{32 \times T_{v}}$. Finally, the Classification Layer, comprising two linear layers with a dropout rate of 0.3, reduces the feature dimension to 2. Softmax applied to the output logits $L \in \mathbb{R}^{T_{v} \times 2}$ indicates speaker activity. The entire module is trained using a standard cross-entropy loss function.

\subsection{Second Stage: Target Speaker Extraction}

The TSE module takes the single-channel mixed speech and the VVAD output $L$ as inputs. The real and imaginary parts of the input mixture signal in the T-F domain are stacked as $Y \in \mathbb{R}^{2 \times T \times F}$, where $T$ and $F$ denote the number of speech frames and frequency bins, respectively. Using the Short-Time Fourier Transform (STFT) with a Hanning window (frame length: 320 samples, frame shift: 160 samples), each second of speech produces 100 frames, while video frames are 25 fps. Thus, one video frame spans four audio frames ($T = 4T_{v}$). To align the video and audio frames, the argmax of the VAD sequence is computed, repeated four times per element, and reshaped to $L^{\prime} \in \mathbb{R}^{1 \times T \times 1}$. The element-wise product of $L^{\prime}$ and $Y$ is concatenated with $Y$, yielding $X \in \mathbb{R}^{4 \times T \times F}$, which serves as input to the subsequent model. This alignment helps transfer video modality features into the speech modality, reducing the gap between the two. Next, we will introduce the components of the TSE module: Encoder, Cross-Narrow Band Backbone, and Decoder.

\subsubsection{Encoder}

To ensure computational efficiency in the TSE module, we use an efficient encoder to downsample the input features $X$, reducing the computational cost of the subsequent Cross-Narrow Band Backbone. We adopted the Encoder architecture from GTCRN \cite{rong2024gtcrn}, modifying the channel dimension from 16 to 64 to balance computational cost and modeling capacity. The downsampled features are denoted as $X^{\prime} \in \mathbb{R}^{64 \times T \times F^\prime}$, where $F^\prime$ is the size of the frequency dimension after downsampling.

\subsubsection{Cross-Narrow Band Backbone}
In recent years, networks based on the Cross-Narrow Band architecture have shown great success in speech enhancement and separation tasks \cite{wang2023tf,quan2024spatialnet}. In this work, we use a Cross-Narrow Band architecture as the backbone of our second-stage network, which includes three components: the Cross-Band Module, Narrow-Band Module, and Chunk Attention Module. The network performs no frequency upsampling or downsampling, and the output is $X^{\prime\prime} \in \mathbb{R}^{64 \times T \times F^\prime}$.

To capture cross-band correlations in the input features, we employ the cross-band module from \cite{quan2024spatialnet}, which consists of two FConv blocks and a full-band linear module for modeling correlations across the entire frequency band. The FConv block includes a LayerNorm, a Conv1D layer with a kernel size of 5 along the frequency dimension, and a PReLU activation function. The Full-Band Linear Module starts with a linear layer followed by a SiLU activation, expanding the channels to $H^{\prime}$, which is 128 in our work. A series of linear layers, where each channel is mapped to an independent linear transformation denoted as $\text{Linear}_{i}$, captures the full-band correlations, as shown in Figure \ref{fig:big-fig}(d). Finally, another linear layer with a SiLU activation restores the channel dimensions to the original size. 

The Narrow-Band Module captures long-term dependencies by processing each frequency independently with shared parameters. It consists of a LayerNorm, a single-layer LSTM with 64 units, and a linear layer with input and output dimensions of 64.

The attention mechanism with causal masking has a time complexity of $O(n^2)$ during real-time processing, which poses challenges for edge device deployment. To address this, we adopt the Chunk Attention approach from \cite{veluri2024look}, which limits the temporal scope of the attention layer, reducing its complexity to linear. The Chunk Attention module architecture is shown in Figure \ref{fig:big-fig}(f). Each projection layer (Proj.) consists of a linear layer followed by PReLU and LayerNorm. Additionally, the model maintains K-Cache and V-Cache buffers, denoted as $C_k$ and $C_v$, with a time length of $L$ frames ($L = 50$ in our work). After concatenating the $K$ and $V$ tensors with the cache along the time axis, we apply an unfold operation with a kernel size of $L$ and a stride of 1 to partition them into independent fixed-size blocks. The attention matrix is then computed for each block by comparing the Key tensor with the single-frame Query tensor corresponding to the last frame in the block.

\subsubsection{Decoder}
The decoder mirrors the encoder, with each Conv block replaced by a deconvolution (DeConv) block. Residual connections are incorporated between each layer of the encoder and the corresponding layer of the decoder. The final layer uses a tanh activation to output the CRM for the target and interfering speakers as a 4-channel tensor $M \in \mathbb{R}^{4 \times T \times F}$. The network is trained with a composite loss function, which combines the Mean Squared Error on the magnitude spectrograms and the negative Scale-Invariant Signal-to-Noise Ratio (SI-SNR) of the reconstructed speech signals. During inference, only the target speaker's CRM is used to reconstruct the enhanced speech.

\section{Experiments}
\subsection{VVAD Module Data Preparation}
Training a robust, frame-level VVAD module is challenging due to two limitations in existing datasets. First, large-scale audio-visual corpora like VoxCeleb2 \cite{Chung18b} are severely class-imbalanced, with speech frames (84.64\%) overwhelming non-speech frames (15.36\%). Second, dedicated VVAD datasets like VVAD-LRS3 \cite{lubitz2021vvad} provide only video-level labels for short clips, which lack the natural pauses found in continuous speech and are thus suboptimal for our frame-level prediction task.

To overcome these issues, we employ a two-stage training strategy. First, we pre-train the VVAD module for 25 epochs on VVAD-LRS3 to learn basic visual speech features. Second, to address the class imbalance and introduce realistic speech-pause dynamics, we fine-tune the module on a custom-synthesized dataset. Using Real3D-Portrait \cite{ye2024real3dportrait} with portrait inputs from CelebV-HQ \cite{zhu2022celebvhq}, we synthesized 15 hours of talking portrait videos. By randomizing the duration and position of speaking segments, we created a well-balanced dataset comprising 59.5\% speech frames and 40.5\% non-speech frames, significantly improving the model's robustness for real-world scenarios.

\subsection{TSE Module Data Preparation}
We generated a diverse training dataset for the TSE module on-the-fly. Clean speech was sourced from the 100-hour and 360-hour subsets of the LibriSpeech \cite{panayotov2015librispeech}, utilizing 1,172 speakers for training and holding out 117 for validation. Background noise was drawn from the DNS Challenge 2020 dataset \cite{reddy2020interspeech}. We simulated varied room acoustics by generating Room Impulse Responses (RIRs) using the Image method. Room dimensions $(L,W)$ were sampled from [3,8] m, with height fixed at 3 m, and reverberation time ($T_{60}$) ranged from 0.1 to 0.6 s.

To better reflect real-world scenarios, we deliberately avoided 100\% overlap between the target and interfering speakers. Each mixture was created with a signal-to-interference ratio (SIR) from [-5,5] dB and a signal-to-noise ratio (SNR) from [0,15] dB. Critically, each sample begins with a segment of only the target or the interferer, providing explicit activation cues for the model.

The ground-truth VAD for the target speaker was generated using the WebRTC VAD package\footnote{https://github.com/wiseman/py-webrtcvad?tab=readme-ov-file} and used as an input cue for training the TSE module. To make the TSE module robust to potential errors from the upstream VVAD module, we implemented a VAD augmentation strategy. This involved simulating two common error types observed in our VVAD module: detection delays and misclassifications (label flipping). By training the TSE module with these intentionally corrupted VAD cues, we significantly enhance its robustness for real-world deployment.

\subsection{Evaluation}
To ensure a comprehensive assessment, we evaluated our system's performance on two distinct test sets. First, for direct and fair comparison with state-of-the-art methods, we used the widely-adopted LRS2-2Mix test set, adhering to the same configuration from \cite{lee2021looking}. Second, to assess performance in more realistic conversational scenarios with sparse overlaps, we constructed a custom test set based on the high-quality FaceStar audio-visual dataset \cite{yang2022audiovisual}. For this set, each sample was mixed with an interfering utterance from the LibriSpeech test-clean set and background noise from the DNS Challenge 2020 \cite{reddy2020interspeech} but unseen during training. The acoustic parameters were randomized to simulate diverse environments, with an overlap ratio of [20\%, 80\%], a $T_{60}$ of [0.1, 0.6] s, an SIR of [-5, 5] dB, and an SNR of [0, 15] dB. This custom dataset was used to evaluate both the standalone VVAD module and the complete 2S-AVTSE system.

The VVAD system achieved an accuracy of 78.46\%, a precision of 87.65\%, and a recall of 83.96\% on FaceStar dataset. In subsequent experiments, the inference results generated by this checkpoint were utilized as input for the second-stage processing.

\subsubsection{Performance on LRS2-2mix}
The performance comparison on the LRS2-2mix test set is presented in Table \ref{lrs2-2mix}. We compare our causal 2S-AVTSE system against the non-causal, state-of-the-art CTCNet \cite{li2024audio} and a lightweight version, CTCNet-mini, which has a computational load comparable to our model. It is important to highlight a fundamental mismatch between our model's design and this benchmark: LRS2-2mix consists of fully overlapping speech, whereas 2S-AVTSE is architected to identify the target speaker using activation cues present in sparsely overlapped speech. To enable our model to function in this setting, we prepended a 2-second, non-overlapping segment of either the target or interfering speaker's voice to each test sample, allowing the system to lock onto the correct speaker.

\begin{table}[htbp]
  \centering

  \caption{Evaluation results on LRS2-2mix dataset.}
  \label{lrs2-2mix}
  \resizebox{\linewidth}{!}{%
  \begin{tabular}{lcccccc}
    \hline
    \textbf{Method} & \textbf{Causal} & \textbf{MACs} & \textbf{Parms} & \textbf{SI-SNR} & \textbf{STOI} & \textbf{PESQ} \\
    \hline
    Unprocessed & - & - & - & 0 & 0.66 & 1.53 \\
    CTCNet & \usym{2717} & 92.56G & 18.34M  & 13.72 & 0.92 & 3.07 \\
    CTCNet-mini & \usym{2717} & 2.26G & 0.54M & 8.92 & 0.86 & 2.17 \\
    2S-AVTSE & \checkmark & 1.89G & 1.36M &6.97 & 0.84 & 2.03 \\
    \hline
  \end{tabular}
  }

\end{table}

As the results show, the end-to-end baselines, which are trained and tested on LRS2-2mix, excel in this in-domain task. This outcome is expected, as our model’s architecture is deliberately optimized for generalization to realistic, sparsely-overlapped scenarios rather than for performance on this specific, artificial benchmark.

\subsubsection{Performance on Realistic and Real-World Data}
The performance of our system on the realistic, sparsely-overlapped FaceStar-Mix test set is detailed in Table \ref{albitation}. In this challenging cross-domain scenario, our proposed 2S-AVTSE achieves an SI-SNR of 7.09 dB, outperforming the large CTCNet model. More strikingly, the lightweight CTCNet-mini, which performed reasonably on the in-domain LRS2-2mix data, suffers a catastrophic performance collapse, with its SI-SNR dropping to -0.59 dB, indicating a complete failure to generalize.

\begin{table}[h!]
\vspace{-0.4cm}
\caption{Evaluation results on our realistic conversational test set (FaceStar-Mix). Our proposed method is shown in bold.}
\centering
\resizebox{\linewidth}{!}{%
\begin{tabular}{cccccc}
\toprule
\textbf{Method}&\textbf{Training} & \textbf{Inference} & \textbf{SI-SNR} & \textbf{STOI} & \textbf{PESQ} \\ \midrule
0&\multicolumn{2}{c}{Unprocessed}& 1.06 & 0.71 & 2.05 \\ \hline
\textbf{1}&\textbf{noised VAD} & \textbf{pred VAD} & \textbf{7.09} & \textbf{0.78} & \textbf{2.41}\\  
2&oracle VAD & pred VAD & 2.88 & 0.67 & 1.90 \\
3&oracle VAD & oracle VAD & 8.69 & 0.81 & 2.54 \\
4&\multicolumn{2}{c}{Audio Only (clean Anchor)}&5.48&0.75&2.28\\
5&\multicolumn{2}{c}{Audio Only (noisy Anchor)}&4.51&0.73&2.17\\
\midrule
6&\multicolumn{2}{c}{CTCNet} & 5.90 & 0.77 & 2.20\\
7&\multicolumn{2}{c}{CTCNet-mini} & -0.59 & 0.60 & 1.54\\
\bottomrule
\end{tabular}%
}
\label{albitation}

\end{table}

\begin{figure}[h]
    \centering
    \includegraphics[width=\linewidth]{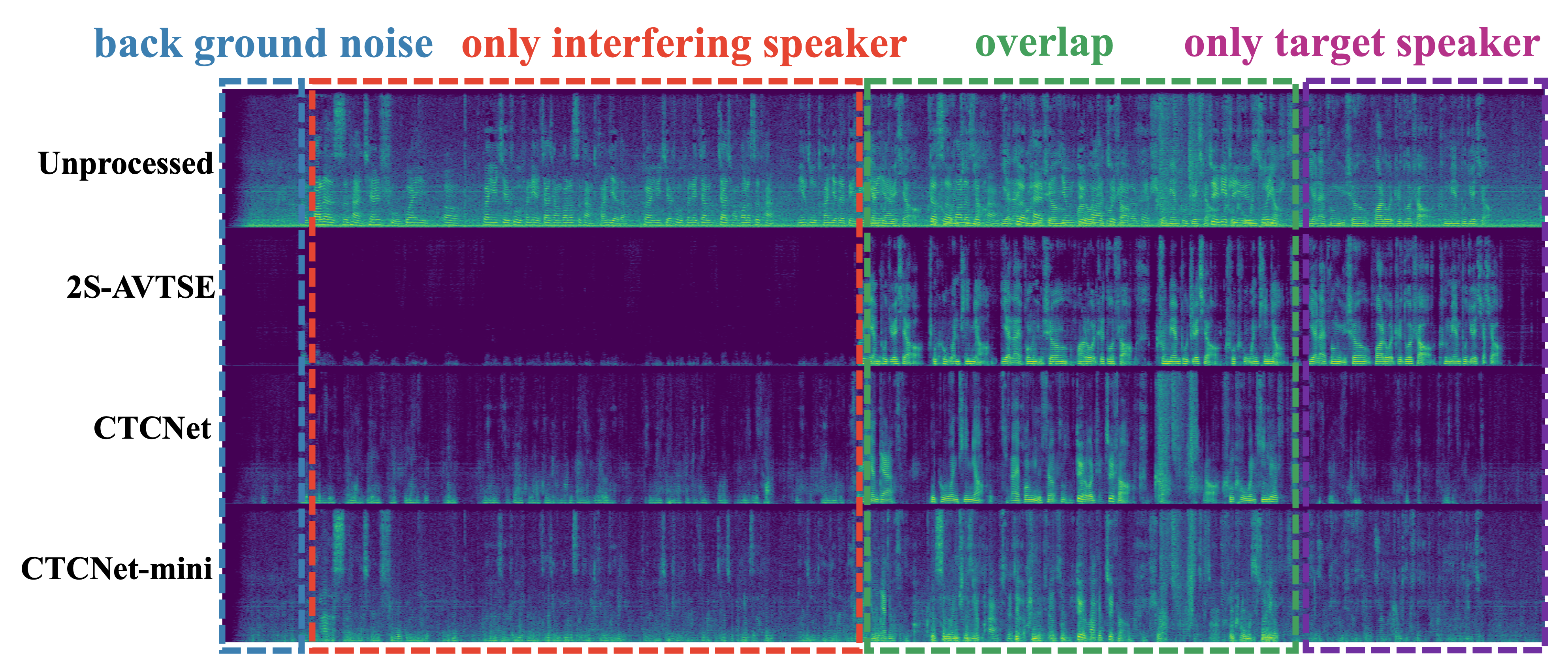}
    \caption{Real recording and the outputs of different methods.}
    \label{fig:demo}
    \vspace{-0.6cm}
\end{figure}

To further validate this in a real-world setting, we recorded a sample\footnote{More demos can be  found in http://www.cslzx.cn/2S-AVTSE/} using a laptop in an office environment, capturing background noise, an interfering speaker, and the target speaker. The resulting spectrograms are shown in Figure \ref{fig:demo}. The visual evidence strongly corroborates our quantitative findings: our 2S-AVTSE system effectively nullifies both background noise and the interfering speaker while preserving the target's speech with high fidelity. While the full CTCNet has over-suppression of the target's voice and residual interference from the non-target speaker. In contrast, CTCNet-mini completely fails to separate the speakers, retaining significant interference. This demonstrates that conventional end-to-end models struggle with real-world generalization, and confirms that our two-stage approach provides a robust, efficient, and truly practical solution.

\subsubsection{Ablation Studies}
The results in Table \ref{albitation} allow for a detailed analysis of our design choices.

\textbf{Importance of VAD Augmentation:} We first validate our VAD augmentation strategy by comparing our proposed method (Method 1) with a model trained on clean VAD labels (Method 2). When subjected to the predicted VAD at inference, Method 2's performance collapses (SI-SNR drops from 7.09 to 2.88 dB). This demonstrates a critical mismatch between its clean training conditions and the imperfect, real-world VAD inputs. This result confirms that our strategy of augmenting VAD labels with noise is essential for making the TSE module robust to errors from the upstream VVAD system.

\textbf{Performance Upper Bound and Future Work:} Method 3 establishes a theoretical performance upper bound by using oracle (ground-truth) VAD labels during inference, achieving an SI-SNR of 8.69 dB. The performance gap between our proposed system (7.09 dB) and this upper bound indicates that the primary bottleneck is the accuracy of the first-stage VVAD module. Therefore, improving the precision of the VVAD system is a clear and promising direction for future work.

\textbf{Comparison with Audio-Only Baselines:} Finally, we compare our visual-cue approach against audio-only baselines (Methods 4 \& 5). To ensure a fair comparison, these baselines utilize the identical speech extraction network as our proposed method, but the guiding cue is derived from a pre-recorded voiceprint (anchor) instead of the visual VAD. Our proposed method significantly outperforms both audio-only variants, proving that for this task, a visual VAD signal is a more effective and robust cue than a spectral embedding from an anchor utterance. Furthermore, our approach carries a significant practical advantage: it eliminates the need for a separate user enrollment step (i.e., pre-recording clean audio), enabling a seamless, "zero-shot" user experience in any environment.

\subsubsection{Ease of deployment}
To assess the real-time performance of the 2S-AVTSE system, we exported the ONNX model using PyTorch 2.1.1 and evaluated its inference time on two typical office laptops: one with an Apple M1 Pro (ARM architecture) and the other with an Intel i5-12450H (x86 architecture). Using the ONNX Runtime (ORT), we performed 1000 consecutive inference operations. The average inference times were 1.46 ms on the M1 Pro and 2.9 ms on the i5-12450H, both comfortably below the 10 ms frame shift required for real-time processing.

\section{Conclusions}
In this paper, we proposed 2S-AVTSE, a two-stage audio-visual TSE framework based on a novel decoupled training paradigm. Our experimental results demonstrate that, in contrast to conventional end-to-end models, 2S-AVTSE achieves superior generalization to realistic, cross-domain scenarios while maintaining a lightweight and causal architecture. These qualities make our system a robust and highly promising solution for practical, real-world deployment.

\textbf{Acknowledgements}: This research was partly supported by the China National Nature Science Foundation (No. 61876214) and CCF-Lenovo Research Fund (No. 20240203).

\bibliographystyle{IEEEbib}
\bibliography{strings,refs}

\end{document}